# Cumulative Sum Algorithm for Detecting SYN Flooding Attacks

Tongguang Zhang

*Department of Computer and Information Engineering*
*Xinxiang College*
*Xinxiang, Henan 453000, China*

jsjoscpu@163.com

*Abstract* – SYN flooding attacks generate enormous packets by a large number of agents and can easily exhaust the computing and communication resources of a victim within a short period of time. In this paper, we propose a lightweight method for detecting SYN flooding attack by non-parametric cumulative sum algorithm. We experiment with real SYN flooding attack data set in order to evaluate our method. The results show that our method can detect SYN flooding attack very well.

## I. INTRODUCTION

Distributed denial of service(DDoS) attacks launched from a large number of compromised hosts, are a major threat to Internet services [1-2]. Popular web sites are among the well known victims of attacks, yet an even larger number of online companies than these depend on the stability and availability of the Internet and face considerable losses should they be the object of a DDoS attack [3-4].

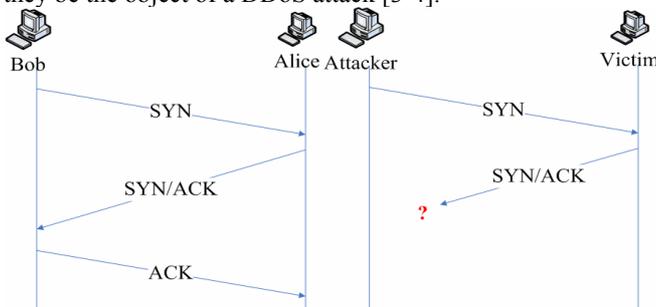

Fig. 1 Normal TCP connection and half-open TCP connection.

In DDOS attacks, more than 90% of attacks exploit TCP. The most efficient and commonly-used SYN flooding attacks exploit the standard TCP three-way handshake in which the server receives a client's SYN request, replies with a SYN/ACK packet and then waits for the client to send the ACK to complete the handshake. The normal behaviour of this process is summarized as follows. If Bod wants to connect to Alice then Bob initiates the connection by sending a SYN segment to Alice. Alice should reply with a SYN/ACK segment that acknowledges the initial SYN. The connection is established between the two endpoints when Alice returns an ACK to Bob, acknowledging the SYN–ACK as shown in the left of Fig. 1. While waiting for the final ACK, the server maintains a half-open connection. Since the SYN flooding attacker always chooses unreachable addresses as the spoofed source addresses of the attacking packets, the server will not receive the anticipated final ACK from the client as shown in the right of Fig. 1. As more and more half-open connections are maintained on a victim server, such SYN flooding attacks deplete the server's resources and the server will as a result refuse services even to legitimate customer requests [5].

At present, many attack detection and defense mechanisms to SYN flooding exist [4, 6-10] . Their goal is to detect, countermeasure, or prevent attacks.

In [6], the authors proposed a novel defense mechanism that makes use of the edge routers that are associated with the spoofed IP addresses' networks to determine whether the incoming SYN–ACK segment is valid. This is accomplished by maintaining a matching table of the outgoing SYNs and incoming SYN–ACKs and also by using the ARP protocol. If an incoming SYN–ACK segment is not valid, the edge router resets the connection at the victim's host, freeing up an entry in the victim's backlog queue, and enabling it to accept other legitimate incoming connection requests. They also presented a communication protocol to encourage collaboration between various networks to protect each other.

In [7] , the authors seek to provide a method that detects SYN flooding attacks in a timely fashion and that responds accurately and independently on the victim side. They used the knowledge of network traffic delay distribution and apply an active probing technique (DARB) to identify half-open connections that, suspiciously, may not arise from normal network congestion. This method is suitable for large network areas and is capable of handling bursts of traffic flowing into a victim server. Accurate filtering is ensured by a counteraction method using IP address and time-to-live (TTL) fields.

In [8], the authors' main motivation for doing so is network security: rapid anomaly detection for an early detection of attacks in computer networks that lead to changes in network traffic. Moreover, this kind of application encourages the development of a nonparametric multichannel detection test that does not use exact pre-change (legitimate) and post-change (attack) traffic models. The proposed nonparametric method can be effectively applied to detect a wide variety of attacks such as denial-of-service attacks, worm-based attacks, port-scanning, and man-in-the-middle attacks. In addition, they also proposed a multichannel procedure that is based on binary quantized data; this procedure turns out to be more efficient than the previous two algorithms in certain scenarios.

In this paper, we propose a lightweight SYN flooding attack detection method. In this method, we adopt non-parametric cumulative sum algorithm to find attacks. We conduct the experiments to evaluate our proposed method. Experimental results show that our method can effectively detect TCP SYN flooding attacks

The remainder of this paper is organized as follows. Section II introduces our proposed method. Experiments are taken in Section III. Finally, Section IV is our conclusions.

## II. OUR METHOD

Cumulative sum (CUSUM) algorithm is used in the quality control. They are well suited for checking a measuring system in operation for any departure from some target or specified values and have been widely used for detecting the small and moderate mean shifts [11-12].

In this paper, we focuses on the use of non-parametric CUSUM [12] to detect TCP SYN flooding attacks. In the context of detecting SYN flooding attacks, for each SYN packet, CUSUM monitors a set of *n* SYN packet sample interval $\{y_1, \cdots, y_n\}$. $y_n$ is the sum of all SYN packets in *n-th* sample interval (detection interval). Assume that the change SYN traffic $\{y_i\}$ is independent Gaussian distribution with known variance $\sigma^2$ [13], which we assume remains the same after the change, and $\mu_0$ and $\mu_1$ are the mean SYN traffic before and after the change.

Then CUSUM ($f_n$) can be described as follows:

$$f_n = \left[ f_{n-1} + \frac{\mu_1 - \mu_0}{\sigma^2}(y_n - \frac{\mu_1 + \mu_0}{2}) \right]^+ \quad (1)$$

Of course the assumption of Gaussian distribution about $\{y_i\}$ can be not true for TCP SYN packet measurements, due to weekly or daily variations, trends, and time correlations. Thus, such non-stationary behavior should be removed before applying CUSUM by the approach in [14]. In addition to leading to complex and time-consuming calculations, we consider a simple approach to apply CUSUM to $\tilde{x}_n$, with

$$\tilde{x}_n = x_n - \bar{\mu}_{n-1} \quad (2)$$

where $x_n$ is the sum of all SYN packets in the *n-th* sample interval, and $\bar{\mu}_n$ is an estimate of the mean number of SYN packet at sample n, which is computed using an exponential weighted moving average (EWMA) as follows:

$$\bar{\mu}_n = \lambda \bar{\mu}_{n-1} + (1-\lambda)x_n \quad (3)$$

where $\lambda$ is the EWMA factor. The mean rating traffic of $\tilde{x}_n$ prior to a change is zero, hence the mean in (2) is $\mu_0 = 0$. A remaining issue that needs to be addressed is the value of $\mu_1$, i.e., the mean number of SYN packet after the change. This cannot be known beforehand, hence we approximate it with $\alpha \bar{\mu}_n$, where $\alpha$ is amplitude percentage parameter, which corresponds to the most probable percentage of increase of the mean number of SYN packet after a change.

Then CUSUM from (1) can be written:

$$f_n = \left[ f_{n-1} + \frac{\alpha \bar{\mu}_{n-1}}{\sigma^2}(x_n - \bar{\mu}_{n-1}\frac{\alpha \bar{\mu}_{n-1}}{2}) \right]^+ \quad (4)$$

If $f_n \geq h$ ($h > 0$ is the predesigned CUSUM threshold parameter), then alarm.

## III. PERFORMANCE EVALUATION

### A. Experiments setup

We carried out a simulation experiment to evaluate the performance of our proposed method. The performance consists of the detection ratio and false alarm ratio.

In this experiment, we adopt a real TCP SYN flooding attacks dataset, i.e., DARPA dataset [15]. It is an off-line intrusion detection benchmark is used for the experiment. The DARPA data set represents Tcpdump and audit data generated over five weeks of simulated network traffic in a hypothetical military local area network (LAN). The DARPA data set was the first in a planned series of annual evaluations conducted by MIT Lincoln Laboratory under DARPA sponsorship [1, 5]. These evaluations are designed to focus research efforts on core technical issues and provide unbiased measurement of current performance levels. The primary purpose of the evaluations is to drive iterative performance improvements in participating systems by revealing strengths and weaknesses and helping researchers focus on eliminating weaknesses. To insure that the greatest numbers of researchers can participate, common shared corpora are created that can be distributed and used by a wide range of researchers [16]. Such corpora simplify entrance into this field and make it possible to compare alternate approaches. To make sure the evaluation could uncover weaknesses in many types of intrusion detection systems, widely varied attacks were developed that span the types of attacks which might be used by both novice and highly skilled attackers. Efforts were also made to keep the evaluation simple and to encourage the widest participation possible by eliminating security and privacy concerns and providing data types used by the majority of intrusion detection systems. Simplicity and more widespread participation were obtained in the first 1998 evaluation by focusing on UNIX hosts and outside attacks originating from remote hosts [15, 17].

### B. Performance result on detection ratio

Table I show the detection results with our proposed method. From the results, our proposed method is very effective. It can accurate detect TCP SYN flooding attacks with 98.82% detection ratio on average. For example, Although when the number of TCP SYN attack packet is smaller than 400, the detection ratio cannot reach to 100%, when the number is bigger than 600, the detection ratio of our proposed method is 100%. This means that our proposed method is effective. It can find the SYN flooding attack with high detection ratio.

Moreover, in this paper, if we cannot consider the dynamic network environment, the experimental results of our proposed method indicate that our proposed method can almost find all TCP SYN flooding attacks.

TABLE I
DETECTION RATIO

| SYN packets/second | Detection ratio |
|---|---|
| 200 | 95.9 % |
| 400 | 98.2 % |
| 600 | 100 % |
| 800 | 100 % |
| 1000 | 100 % |

*C. Performance result on false alarm ratio*

From the Table 2, our proposed approach is also very effective. Its false alarm ratio is very low on average. It only is 2.46%. The false alarm ratio is so low that it almost is omitted in practical application.

Furthermore, When the number of TCP SYN attack packet is bigger than 800, the false alarm ratio is 0%. This means that our proposed method accurately can find SYN flooding attacks. Although when the number is smaller than 600, the false alarm ratio of our proposed method isn't 0%, it also is smaller that 10%. Moreover, in this paper, if we cannot consider the dynamic network environment, the experimental results of our proposed method indicate that our proposed method can almost find all TCP SYN flooding attacks.

From Tables I and II, the experimental results indicate that our proposed method is very effective. It accurately can find TCP SYN flooding attack with low false alarm ratio and high detection ratio.

TABLE II
FALSE ALARM RATIO

| SYN packets/second | False alarm ratio |
|---|---|
| 200 | 7.4 % |
| 400 | 4.3 % |
| 600 | 0.6 % |
| 800 | 0 % |
| 1000 | 0 % |

IV. CONCLUSIONS AND FUTURE WORK

TCP SYN flooding attacks seriously threaten Internet services yet there is currently no defense against such attacks that provides both early detection, allowing time for counteraction, and an accurate response. Traditional detection methods rely on passively sniffing an attacking signature and are inaccurate in the early stages of an attack. Current counteractions such as traffic filter or rate-limit methods do not accurately distinguish between legitimate and illegitimate traffic and are difficult to deploy. This work seeks to provide a lightweight method that accurately can find SYN flooding attacks. In this paper, we apply non-parametric cumulative sum algorithm to detect SYN flooding attack. In order to evaluate our proposed method, we conducted experiment with a real TCP SYN flooding attack dataset. The experimental results shown that our proposed method is very effective. It accurately can find TCP SYN flooding attack with low false alarm ratio and high detection ratio.

Although our proposed method is very effective, there are some limitations in our method. For example, how to evaluate the detection time of our proposed method. Hence, in order to improve the performance of our method, it is our future research work.